# S-Mbank: Secure Mobile Banking Authentication Scheme Using Signcryption, Pair Based Text Authentication, and Contactless Smartcard

Dea Saka Kurnia Putra
Manajemen Persandian
Sekolah Tinggi Sandi Negara
Bogor, Indonesia
dea.saka@student.stsn-nci.ac.id

Mohamad Ali Sadikin
Teknik Persandian
Sekolah Tinggi Sandi Negara
Bogor, Indonesia
mohamad.ali@student.stsn-nci.ac.id

Susila Windarta
Sekolah Tinggi Sandi Negara
Bogor, Indonesia
susila.windarta@stsn-nci.ac.id

*Abstract*— Nowadays, mobile banking become a popular tool which consumers can conduct financial transactions such as shopping, monitoring accounts balance, transferring funds and other payments. Consumers dependency on mobile needs, make people take a little bit more interest on mobile banking. The use of one time password which is sent to the user mobile phone by short message service (SMS) is a vulnerability which we want to solve with proposing a new scheme called S-Mbank. We replace the authentication using one time password with contactless smartcard to prevent attackers to use the unencrypted message which is sent to the user's mobile phone. Moreover, it deals vulnerability of spoofer to send an SMS pretending as a bank's server. Contactless smartcard is proposed because of its flexibility and security which easier to bring in our wallet than the common passcode generators. The replacement of SMS-based authentication with contactless smartcard removes the vulnerability of unauthorized users to act as a legitimate user to exploit the mobile banking user's account. Besides that, we use public-private key pair and PIN to provide two factors authentication and mutual authentication. We use signcryption scheme to provide the efficiency of the computation. Pair based text authentication is also proposed for the login process as a solution of shoulder-surfing attack. We use Scyther tool to analyze the security of authentication protocol in S-Mbank scheme. From the proposed scheme, we are able to provide more security protection for mobile banking service.

*Keywords— Secure mobile banking; Two factor authentication; Mutual authentication; Contactless smartcard; Signcryption scheme; Pair based text authentication; Mobile phone; Scyther Tools;*

## I. INTRODUCTION

Mobile phones nowadays become tools that consumers use for budgeting, shopping, banking, and payments. Based on Board of Governors of the Federal Reserve System reports on March 2016, the use of mobile banking system keep rising [1]. About 43% of mobile phone owners with a bank account had used mobile banking system, it raises from 39% in 2014 [1]. Also, fifty-three percent of smartphone owners with a bank account had used mobile banking, up from 52% in 2014. Among all mobile phone users, 24% believed that people's personal information is "somewhat unsafe" when they use mobile banking, and 18% believed that it is "very unsafe" [4]. Another 15% of mobile phone users simply did not know how safe it is to use mobile banking. And finally, only 8% said it was "very safe" to use mobile banking.

Currently most of online banking services are still applying the primitive system of single factor authentication based solely on the confidentiality of the user ID and PIN [2]. Passwords used on the scheme can be text-based login (textual password) or image-based (graphical password). Text-based passwords are vulnerable to dictionary attack attempts, brute force attacks and shoulder surfing. While image-based passwords require greater resources and processing time is slower. In addition, image-based passwords are vulnerable to shoulder surfing.

Extra strength provided by banks is by generating random values (OTP) by token [3]. Authentication techniques such as single factor authentication cannot avoid direct attacks such as phishing, malware and in-the-middle/browser (MITM) [1], or a type of attack today is the synchronization token [4]. One of the vulnerabilities that arise are when customers access the E-banking services by using public internet access as in hotels, airports, cafes or eating places which unbeknown security is guaranteed. In addition to attacks using malware unwittingly steal confidential information of his customers. Moreover, the attacks on random value using the token is becoming very vulnerable to attack. The latest attacks were successfully carried out, namely the technique of phishing and malware (also called synchronization token) by pretending to ask for customer input response generated by the token to guess the value of the next generation is done [4].

Today many developing authentication method that is able to ensure the confidentiality of customer's data. One of them is to apply the method PBTA [18] [19], mutual authentication, two factor authentication and contactless smartcard. PBTA used to secure the login process from shoulder surfing attacks. Mutual authentication is to ensure entity in internet banking process. Two factor authentication ensures security against direct attacks such as phishing and man in the middle attack while the contactless smartcard [23] to apply the principles of smart authentication that ensures the speed of the authentication, in addition to the contactless smartcard also provides security storage of encrypted private key.

Most of proposed solution has their own advantages and disadvantages. On the other side, all of the solution only authenticates the entities. So that, the needs of mutual authentication for both sides of the bank and the user, so the mobile banking transaction become more secure. We use signcryption scheme to make efficiency in computation than sign-then-encrypt scheme [26] [22]. Signcryption scheme combines sign and encrypt in one process.

## II. BACKGROUND AND RELATED WORKS

Many security risk can be found on computing device especially in mobile device has some different risk on it, below are the security risk can be found in mobile devices [25].

1. Smaller form factor which causes mobile device is more susceptible to theft or loss.

2. More tendency for user to use device in a more personal and confidential way.

3. Unbounded limitation of processing power and battery life which causes many security risks.

According to [25], mobile device has many complexes risk related to the rapid growth of technology, i.e. the use of malware to obtain information illegally which causes privacy violations, wireless carrier infrastructure vulnerabilities, payment infrastructure threats, vulnerabilities in SMS system, hardware and operation system vulnerabilities, complexity of supply chain and new entrants of mobile ecosystem.

Nowadays, malware becomes an awesome challenge to the world of mobile device. According to [16], the percentage of malware which is targeted to attack mobile device rises by 155% in the past year. Spyware and Trojan in SMS ecosystem has a very high growth that is 95%. From 400 identified cases in June 2011, the malware problems found increased to 13000 cases in December 2011 [2].

User's privacy is now a challenging issue for mobile device which lately more personalized and related to the user's identity than a personal computer. Legitimate applications sometimes utilized to pass user data to another applications or used in an unauthorized manner is now catching more attention. Practically problems can be easily found around us, for example given that a case related to European customer's privacy which is sent to United Stated based advertiser. From this case, we found that several applications were accused in breaking European laws of data protection by passing the customer's privacy to United States advertising firm named MobClix without obtaining explicit permission from the users [25].

From the side of internet, mobile devices have many risk related to the network of mobile communications. Wireless carrier is the primary interface of mobile device, especially the radio component called cell sites. The cell sites later communicate through dedicated circuits or microwave to the Mobile Switching Center also called MSC which contains voice and data processing equipment and systems. MSC contains gateway to the internet and another carrier networks. This part usually utilized to get customer's data.

From the risks above, we can conclude that SMS-based authentication has many risks for the second authentication. We assume that the first authentication factor which uses username and password is secure. Then, to subtitute the SMS-based authentication which is used by bank's mobile banking service we propose to use a smartcard to provide the second authentication method we propose to compete the security services. Based on the SANS Institute InfoSec Reading Room paper, the smartcard option is categorized to the forth category, as random number created by an electronic device other than the e-token.

Near Field Communication is used to implement the uses of smartcard in authenticating between user and server [9]. The NFC is using the standardized protocol defined in ECMA-240: 2013 which is compliant to ISO/IEC 18092 standard for Near Field Communication Interface and Protocol 1 (NFCIP-1). NFCIP-1 using inductive coupled devices operating at the centre of frequency of 13,56 Mhz for interconnection of computer peripherals for its communication modes. NFCIP-1 is specified for simple wireless communication between coupled devices with transfer rates of 106, 212, and 424 kbps.

Based on FFIEC IT Examination Handbook, there are some SMS-based authentication risks, such SMS typically transmitted unencrypted over the telecommunications networks and vulnerable to spoofing which allows an unauthorized user to send an SMS pretending to be from a different mobile number to mislead a customer into providing sensitive information to the unauthorized user. Fraudulent SMS messages may also mislead customers into revealing financial institution account information or information used to access financial systems, even the smartcard also have the probability to be stolen or misplaced which may allow unauthorized access to the mobile wallet or user credentials [8].

## III. S-MBANK: OUR PROPOSED SCHEME

In this research, authentication scheme that we propose to develop mobile banking that ensure security of customer data. We use singcryption scheme to make efficientcy in computation [26] [22]. S-MBank uses three additional elements to construct the scheme, which are:

1) PIN

2) Public Key Encryption (PKE) key pairs

3) Smartcard

The key pair will be the second authentication's factor and also provides mutual authentication service [6]. Smartcard is the third factor of authentication's factor and the PIN will be the fourth authentication's factor. Software will be installed on smartphone and smartcard will easily facilitate user to perform authentication process. User will also use Pair Based Text Authentication to enter the password login. PBTA use to prevent shoulder surfing of password [18][19]. Fig 1. Illustrates the proposed scheme. We use the algorithm of Fisheryates Random Number Generator to randomized the pair based text box [9]. Contactless smartcard is also used in the second factor to provide the second factor of entity and data authentication to replace the one time password which is sent by SMS.

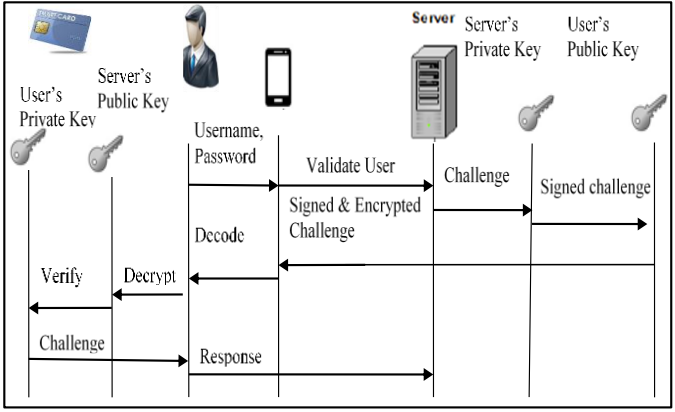

Fig. 1. The proposed Secure Mobile Banking Authentication Scheme

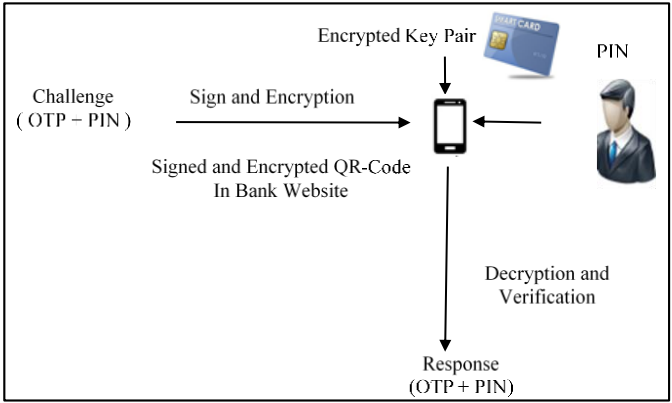

Fig. 2. Work flow for proposed authentication scheme

According to Fig. 1 and Fig. 2, Table I shows the notations and definitions used in S-MBank scheme.

TABLE I. NOTATION AND DEFINITION

| Notation | Definition |
|---|---|
| $U$ | User |
| $S$ | Bank Server |
| $P$ | User's phone number |
| $a$ | Username |
| $b$ | Password (Pair Based Text) |
| $R_r$ | Response nonce |
| $R_c$ | Challenge nonce |
| $K_v$ | Private key |
| $K_p$ | Public key |
| $f$ | Function to generate nonce |

To explain the Figure 1, detailed protocol and its explanation is showed below:

1. User enters username normally and password using Pair Based Text Authentication to prevent shoulder surfing, brute force, and dictionary attack. Server respond it with verifying the username and password by comparing it with database. If the username and password is entered correctly, it will proceed to the next step. Otherwise, it will redirects to the login page.

$$U \rightarrow S : a, b$$

2. Server generates nonce namely Challenge nonce. This Challenge nonce will be combined with user's handphone number to compute Response nonce and save it in a database.

$$S : R_r = f(R_c \| P)$$

3. Server signcrypting the nonce by signed hash value of nonce with the bank's private key and a half part of user's public key. Then server encrypts the nonce with the other a half part of user's public key.

$$U \leftarrow S : KP_U(KV_S(H(R_c)))$$

4. The server send the signcrypted nonce and display it to user on the mobile banking application.

5. The user places the smart card nearby smartphone and convert back the code into signcrypted nonce.

$$U : KP_U(KV_S(H(R_c)))$$

6. User unsigncrypts nonce with the private key of user and the public key of bank.

$$U : KP_S(KV_U(KP_U(KV_S(R_c))))$$
$$U : KP_S(KV_S(R_c))$$
$$U : R_c$$

7. User computes the combination of nonce, and user's handphone number itself using the mobile banking application to get the Response nonce.

$$U : R_r = f(R_c \| P)$$

8. The user sends the "response" into the bank's server.

$$U \rightarrow S : R_r$$

9. Server verifies Response nonce from the user, if it is identical, then the user can perform the further banking transaction. Otherwise, the authentication process terminated.

IV. ANALYSIS RESULTS USING SCYTHER TOOL

Scyther is a tool which was released in 2006 by Casimier Joseph Franciscus Cremers from his thesis titled "Scyther – Semantics and Verification of Security Protocols". Scyther tool is formal automated protocol analysis tool in cryptography protocols using perfect cryptography assumption which assumes every of cryptographic function is perfect. Scyther allows user to compile multiple protocols in a run, which is the advantage of scyther tool than the other protocol analyzers. Adversaries cannot understand anything from encrypted messages except they know the decryption keys. This tool can be used to find the problems from how the protocol is constructed. This problem generally having a characteristic of undecidable, but practically many protocols can be proven whether it is correct or any attacks can be found. To show the vulnerabilities appear in the protocol, we provide the analysis results from the Scyther tool.

TABLE II. COMPARISON OF AVAILABILITY AND FALSIFICATION PROVIDED BY AUTOMATIC VERIFICATION TOOLS

| Verification tool | Public availability | Falsification |
|---|---|---|
| NRL | Not available | No |
| Athena | Not available | Yes |
| LySatool | Available | No |
| Constraint solver | Available | Yes |
| OFMC | Available | Yes |
| Scyther | Available | Yes |

TABLE III. COMPARISON OF VERIFICATION METHOD AND TERMINATION PROVIDED BY AUTOMATIC VERIFICATION TOOLS

| Verification tool | Verification method provided | Termination |
|---|---|---|
| NRL | Unbounded | Yes |
| Athena | Bounded and Unbounded | Yes and No |
| LySatool | Unbounded | Yes |
| Constraint solver | Bounded | Yes |
| OFMC | Bounded | Yes |
| Scyther | Bounded and Unbounded | Yes |

Table 2 and Table 3 shows that every automatic verification tool have their own characteristics. Most of verification tool is easy to get publicly, such LySatool, Constraint solver, OFMC, and Scyther. Because of the newness of Scyther, it is desirable to choose Scyther as the verification tool used to analysis a protocol. Scyther also have its own advantage which provides falsification on the protocol, bounded and unbounded verification method, and termination. The availability of falsification service means that the tool can find attacks which can be useful to understands or repairs a protocol. Falsification also shows the nature of the problem relies on the protocol. Scyther provides bounded verification method means that Scyther typically do a model-checking based approaches into the protocol. On the other sides, unbounded verification method means that Scyther provides a service that we can establish our own properties hold within the security of the protocol. Termination availability indicates whether the verification process of the automatic verification tool is guaranteed to terminate properly.

Table 4 show the source code in Scyther that contain of three roles.

TABLE IV. THE SOURCE CODE OF S-MBANK PROTOCOL

```
protocol SEMBank-Authentication(A,Server)
{role A  {
     fresh PW: Ticket;
     fresh P: Ticket;
     fresh Nr: Nonce;
     fresh Nr1: Nonce;
     var Nr3:Nonce;
     var Nr2:Nonce;
     fresh kpc:SessionKey;
     send_1(A,Server,W,Z);
     recv_2(Server,A,X,Nr3);
     send_3(A,Server,AuthU);
     claim_i1(A, Secret,PW);
     claim_i2(A, Secret, Nr);
          claim_i3(A, Secret, Nr1);
               claim_i4(A, Secret, Nr2);
               claim_i5(A, Secret, Nr3);
               claim_i6(A,SKR,kpc);
               claim_i7(A,Nisynch);
     claim_i8(A,Niagree);   }
  role Server
  {
     fresh P: Ticket;
     fresh Nr1: Nonce;
     fresh Nr3:Nonce;
     fresh Nr2:Nonce;
     var Nr: Nonce;
     var PW: Ticket;
     var kpc:SessionKey;
     recv_1(A,Server,W,Z);
     send_2(Server,A,X,Nr3);
          recv_3(A,Server,AuthU);
          claim_i1(Server, Secret,PW);
          claim_i2(Server, Secret, Nr);
          claim_i3(Server, Secret, Nr1);
          claim_i4(Server, Secret, Nr2);
          claim_i5(Server, Secret, Nr3);
          claim_i6(Server,SKR,kpc);
          claim_i7(Server,Nisynch);
          claim_i8(Server,Niagree);  }
}
```

Fig. 3. The Scyther Verification Result

From the verification result above, we have the fact that our proposed scheme has no vulnerability within bounds which the Scyther verify. It means that our proposed scheme has no possibility of attacks.

V. CONCLUSION AND FUTURE WORK

The design of secure mobile banking scheme we proposed is applying the pair based text authentication method to input the password which the users input onto the login page which has the goal to improve the protection of the vulnerability in shoulder-surfing attack usually done in mobile devices from people around us. This scheme uses Fisheryates Random Number Generator algorithm and replace one time password with contactless smartcard to provide the data and entity authentication and also because the prospective of the near field communication technology in the future. This replacement is stimulated by the messages sent by SMS is trasmitted unencrypted or in plaintext, so that very vulnerable to the attack of unauthorized people. We use ECSCS1 because of its advantage in implementation which has 58% lower in average time of computation. It also has a better memory size usage which is 70% lower than signature-then-encryption scheme in the discrete logarithm problems. Compared with the usage of signature-then-encruption with the RSA algorithm system, the signcryption scheme needs 50% lower computation's time in average and 91% lower in message space. From the automated analysis using Scyther, the protocol has no vulnerability yet.

ACKNOWLEDGMENT

This research was financially and non financially supported by Sekolah Tinggi Sandi Negara. We thank our colleagues from Sekolah Tinggi Sandi Negara who provided insight and expertise that greatly assisted the research, although they may not agree with all of the interpretations/conclusions of this paper.

REFERENCES

[1] Aggarwal, Sunil. 2014. Handbook on Mobile Banking: Regulatory and On-Boarding Guidelines for Regional Rural Banks and Central Cooperative Banks. New Delhi, India.

[2] Aloul, et al. 2009. Two Factor Authentication Using Mobile Phones. ACS International Conference on Computer Systems and Applications (AICCSA). Pp 641-644. Morocco.

[3] Bendersky, E., 2010. Eli Bendersky's website. [Online] Url : http://eli.thegreenplace.net/ [Diakses 27 November 2016].

[4] Board of Governors of the Federal Reserve System. 2016. Consumers and Mobile Financial Services 2016.

[5] C. Cremers. 2014. Scyther User Manual, 2014.

[6] Cooper, et al. RFC 5280: Internet X-509 Public Key Infrastructure Certificate and Certificate Revocation List (CRL) Profile. Internet Engineering Task Force, 2008.

[7] Fang X., and J. Zhan, "Online Banking Authentication Using Mobile Phones," International Conference on Future Information Technology, pp. 1-5, 2010

[8] Federal Financial Institution Examination Council (FFIEC). 2016. FFIEC Information Technology Examination Handbook Appendix E: Mobile Financial Services

[9] Fisher, R. A. & Yates, F., 1963. Statistical Tables For Biological, Agricultural, and Medical Research. 6th penyunt. London: Hafner Press.

[10] Gandhi, A., et.al., 2014. Advanced Online Banking Authentication System Using One Time Passwords Embedded in Q-R Code. International Journal of Computer Science and Information Technologies, Vol.5 (2), 2014, pp 1327-1329.

[11] Gummesson, E., & Lovelock, C. (2004).. Whither Services Marketing? : In Search of a New Paradigm and Fresh Perspectives Journal of Service Research, 7(1). 20-41.

[12] Hiltgen A., T. Kramp and T. Wigold, "Secure Internet Banking," IEEE Security & Privacy, vol. 4, pp. 21-29, 2006.

[13] International Finance Corporation World Bank Group. 2010. Mobile Banking in Indonesia: Assessing the Market Potential for Mobile Technology to Extend Banking to the Unbanked and Underbanked Final Report.

[14] Kumar, S., Temkar, R., Raj, N., 2013. QR Code Based Secure OTP Distribution Scheme for Authentication in Net-Banking. International Journal of Information Science and Intelligent System, Vol.2, Issue 4, pp 115-121.

[15] Liu, Fenghui. 2013. Efficient Two-Factor Authentication Protocol Using Password and Smart Card. Journal of Computers, Vol. 8, No. 12, Desember 2013. Pp 3257-3263. Academy Publisher.

[16] Munandar, M. & Hakim, A. R., 2013. Analisis Keamanan pada Based Text Authentication pada Skema Login. Surabaya: Seminar Nasional Sistem Informasi Indonesia.

[17] Muzakki, K., 2015. Dana Nasabah Rp 130 M Dibobol. Koran SINDO, 16 April. 1. 2b.

[18] Nurchahyanto, Eko Septian. 2015. Penerapan Metode Pair Based Text Authentication Scheme pada Aplikasi Screen Lock sebagai Alternatif dalam Meningkatkan Keamanan Smartphone Android (Ice Cream Sandwich). Bogor: Sekolah Tinggi Sandi Negara

[19] Otoritas Jasa Keuangan(OJK). 2015. Indonesian Banking Statistics. Vol: 13 No. 11, Oktober 2015.

[20] PwC Indonesia. 2015. Indonesian Banking Survey 2015.
[21] Rustandi, Dedy. Smart Card [online]
[22] Smith, C. (2005). Digital Signcryption. University of Waterloo.
[23] Stallings, W. 2014. Crytography and Network Security : Principles and Practices 6th ed. Pearson Education Inc.
[24] Waldman, Jon. 2016. Secure Banking Solution (SBS) Institute. FFIEC Guidance: Mobile Financial Services.
[25] Weerasinghe et al. 2012. Security Framework for Mobile Banking.
[26] Zheng, Y. 1997. Digital Signcryption or How to Achieve Cost (signature & encryption) << Cost (signature) + Cost (encryption). Monash University.